\documentclass[aps,showpacs,pre,twocolumn]{revtex4-1}
\usepackage[T1]{fontenc}
\usepackage[latin1]{inputenc}
\usepackage{amsmath} 
\usepackage{amsthm} 
\usepackage{amssymb}	
\usepackage{graphicx}
\usepackage{epstopdf}
\usepackage{color}
\usepackage[colorlinks=true,linkcolor=blue,citecolor=blue]{hyperref}  
\usepackage{extramarks}
\usepackage{ae}

\begin{document}
\title{Peregrine comb: multiple compression points for Peregrine rogue waves in periodically modulated nonlinear Schr\"odinger equations}
\author{C. G. L. Tiofack, S. Coulibaly, and M. Taki}

\affiliation{Laboratoire de Physique des Lasers, Atomes et Molécules, CNRS UMR
8523, Université Lille 1 - 59655 Villeneuve d'Ascq Cedex, France}

\author{S. De Bi\`evre and G. Dujardin }
\affiliation{Laboratoire Paul Painlev\'e, CNRS,
Universit\'e Lille 1; \'Equipe-Projet Mephysto, INRIA Lille-Nord Europe, France}

\begin{abstract}
It is shown that sufficiently large periodic modulations in the coefficients of a nonlinear Schr\"odinger equation can drastically impact the spatial shape of the Peregrine soliton solutions: they can develop multiple compression points of the same amplitude, rather than only a single one, as in the spatially homogeneous focusing nonlinear Schr\"odinger equation. The additional compression points are generated in pairs forming a comb-like structure. The number of additional pairs depends on the amplitude of the modulation but not on its wavelength, which controls their separation distance. The dynamics and characteristics of these generalized Peregrine soliton are analytically described in the case of a completely integrable modulation. A numerical investigation shows that their main properties persist in nonintegrable situations, where no exact analytical expression of the generalized Peregrine soliton is available. Our predictions are in good agreement with numerical findings for an interesting specific case of an experimentally realizable periodically dispersion modulated photonic crystal fiber. Our results therefore pave the way for the experimental control and manipulation of the formation of generalized Peregrine rogue waves in the wide class of physical systems modeled by the nonlinear Schr\"odinger equation.  
\end{abstract}
\maketitle

\section{ Introduction}
Rogue waves (RW) originally designate abnormally high oceanic surface waves. These devastating walls of water are short-lived and extremely rare, and constitute one of the fascinating manifestations of the strength of nature: ``they appear from nowhere and disappear without a trace''~\cite{Akhmediev2009}. They have been recognized by sailors for nearly a century, but large scale, systematic scientific studies have been undertaken only in the past twenty years, initially in the field of oceanic waves \cite{Dysthe2009}. 
Experimental evidence of optical RW in a fiber system has been established first in the pulsed regime in 
\cite{Solli2007}, and later in the continuous  wave (CW) regime \cite{Mussot2009}. It has received considerable attention since \cite{Hammani2008,Dudley2008,Dudley}.  Moreover, intensive studies of RW have recently been pursued in diverse fields, e.g. acoustics, capillary and plasma waves, generally under the category of ``extreme and rare events in physics" \cite{Kourakis, Onorato2013}. From a theoretical point of view, the most important progress concerns the nonlinear 
Schr\"odinger equation (NLSE) since it appears as a generic model in diverse nonlinear systems such as surface waves or optical fiber systems. Although there exist different classes of explicit nonlinear solutions to the NLSE, the most important solutions commonly accepted to represent rogue waves are the Akhmediev breathers (AB) and the Peregrine solitons (PS) \cite{Usama}. It has been shown that these solutions show indeed the main signatures of RW and extensive studies are devoted to characterize and control the formation of these solutions \cite{Shrira2010,Kharif}. This includes collisions of AB \cite{Frisquet} or the optimum conditions for the experimental generation of PS \cite{Kibler}. 
Our goal here is to study the evolution and properties of the generalized Peregrine rogue waves (PRWs) in systems described by the NLSE with spatially periodic coefficients. We will limit ourselves here to sinusoidally varying coefficients, which have found an experimental realisation in~\cite{Drokes1, Drokes2}. But other periodic modulations are possible and have been recently realized experimentally~\cite{Rota}. We expect the richness of the nonlinear dynamics of PRWs and the flexibility to manipulate them in optical fiber systems with periodic dispersion to be immensely greater than those for an optical fiber with constant coefficients. 

Our models depend on two parameters: the wavelength and the amplitude of the periodic modulation. We address here the following question. How do the main characteristics of the generalized PRW depend on those two control parameters?
We will in particular see that, depending on the amplitude of the modulation, the generalized PRW can have one or several compression points, whose spatial separation is monitored by the wavelength of the dispersion modulation. We show that, under strong modulation, a comblike structure is formed, that we will refer to as the Peregrine comb. 
We take advantage of the
analytical expressions that can be obtained for the generalized PRW under suitable integrability conditions on the coefficients of the NLSE. 

The paper is organized as follows. In section II we show that  under a suitable integrability condition, the periodic modulation of the coefficients of the NLSE can lead to a generalized AB or  Peregrine solution in which not one, but a finite number of high intensity peaks develop. In section III we will show that the generalized Peregrine solution with  a sinusoidal modulation develops multiple compression points depending on the strength of the modulation, leading to a comblike spatial structure.  
Further properties of the generalized Peregrine solution are studied in section IV. In section V we numerically show that such multiple compression points can also arise in the nonintegrable case, provided the group velocity dispersion (GVD) of the equation changes sign repeatedly. We then further illustrate the phenomenon in the case where the GVD and nonlinear coefficient in the equation have experimentally realistic values for photonic crystal fibers. Section~VI contains our concluding remarks. 

\section{Integrable model: generalized Peregrine solution}\label{s:peregrine}
We consider the NLSE in the form
\begin{eqnarray}\label{eq:nlse}
i\frac{\partial q}{\partial z}-\frac{D_{2}(z)}{2}\frac{\partial^{2}
q}{\partial t^{2}}+R(z)|q|^{2}q=0, \label{1}
\end{eqnarray}
where $D_2(z)$ and $R(z)$ are both taken to be periodic functions of their argument.  In optical fiber systems the above equation describes the dynamics of  the envelope $q(z,t)$ of the electric field~\cite{Agrawal2008,Serkin2000} where $z$ and $t$ are the dimensionless propagation distance along the fiber (measured in units of nonlinear length) and dimensionless time, respectively. The function $D_{2}(z)$ represents the GVD coefficient, and $R(z)$ is the nonlinear one. 

\begin{figure}[t]
\includegraphics[width=4cm,height=2cm]{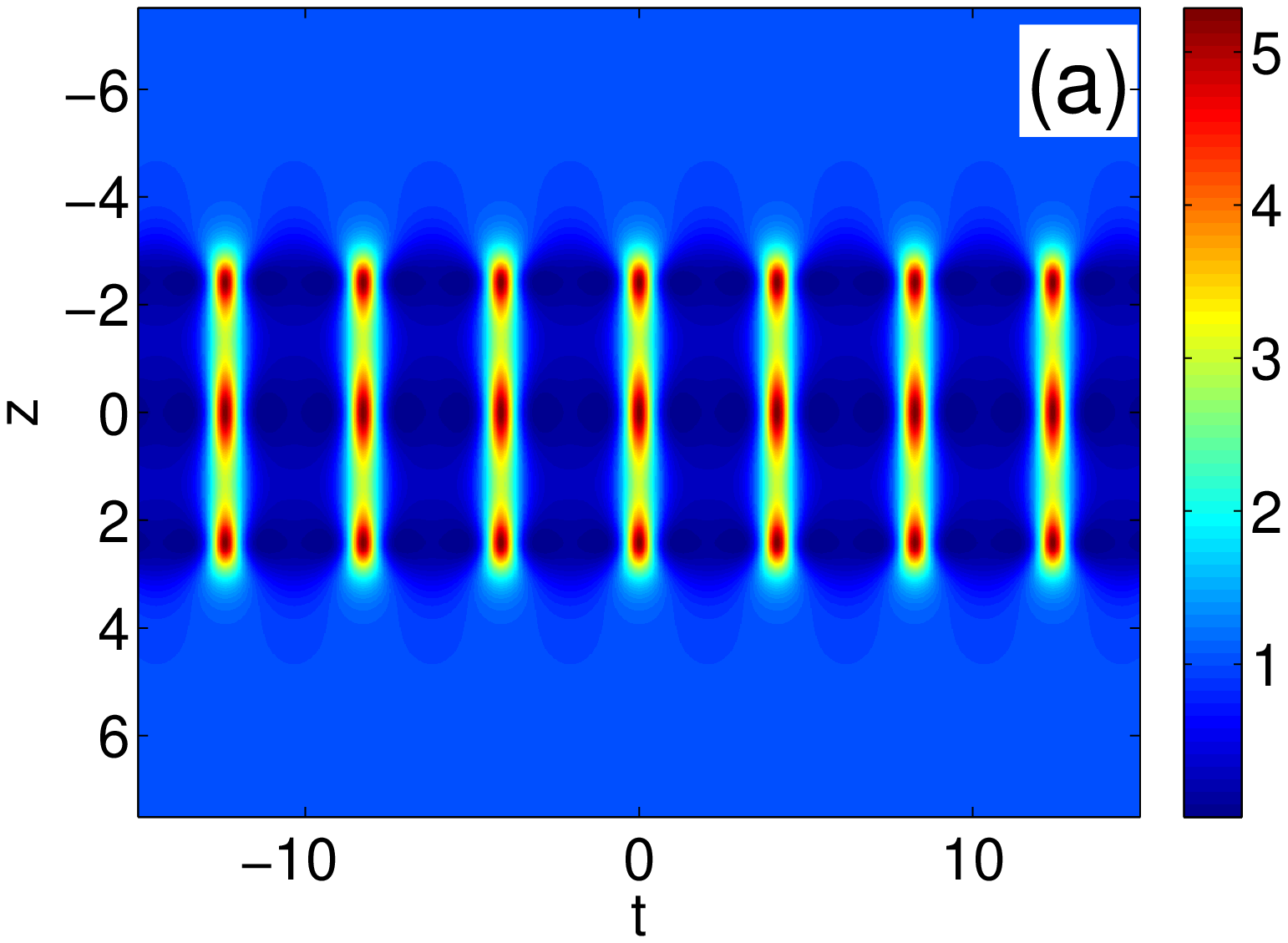}
\includegraphics[width=4cm,height=2cm]{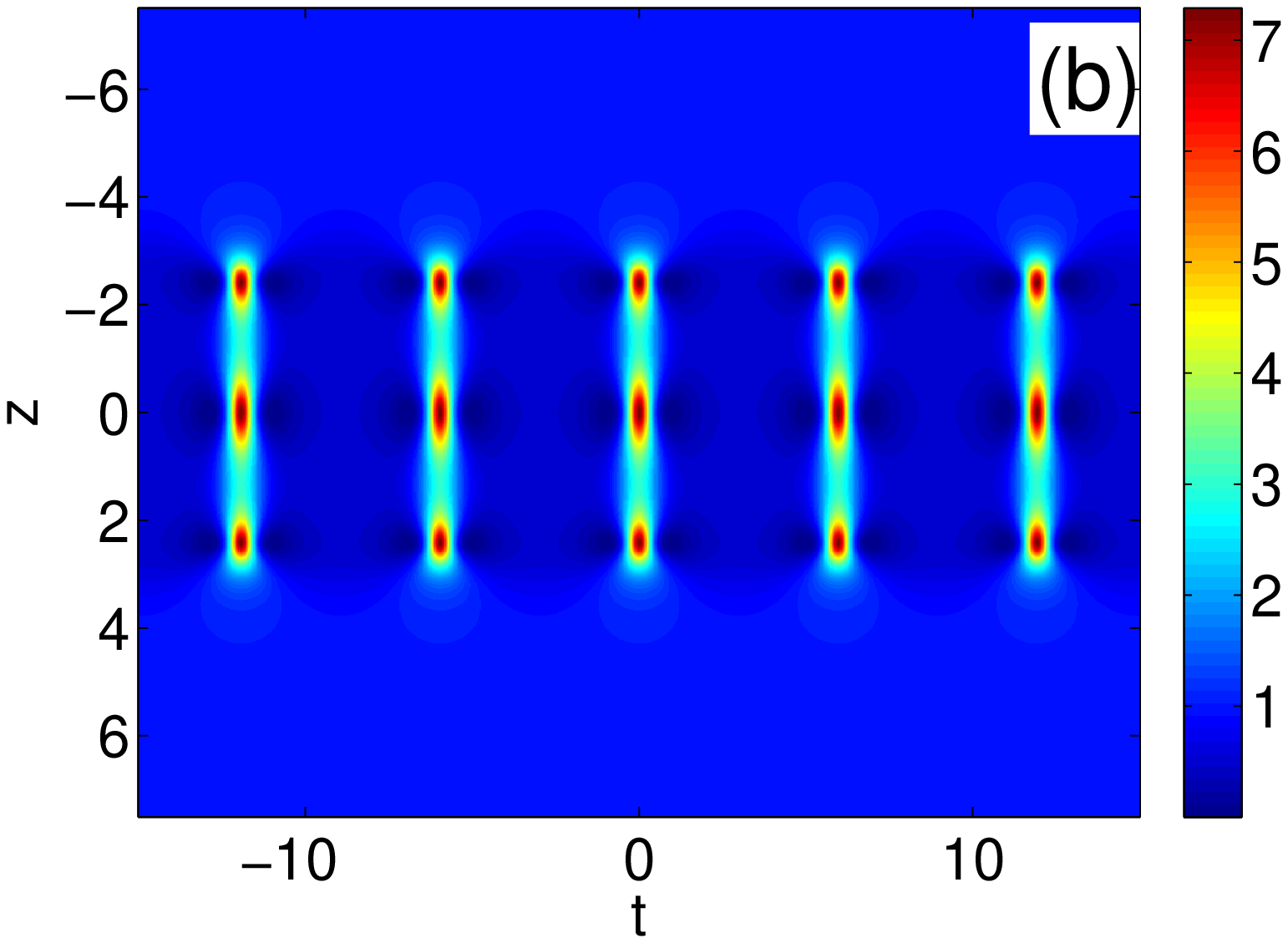}
\includegraphics[width=4cm,height=2cm]{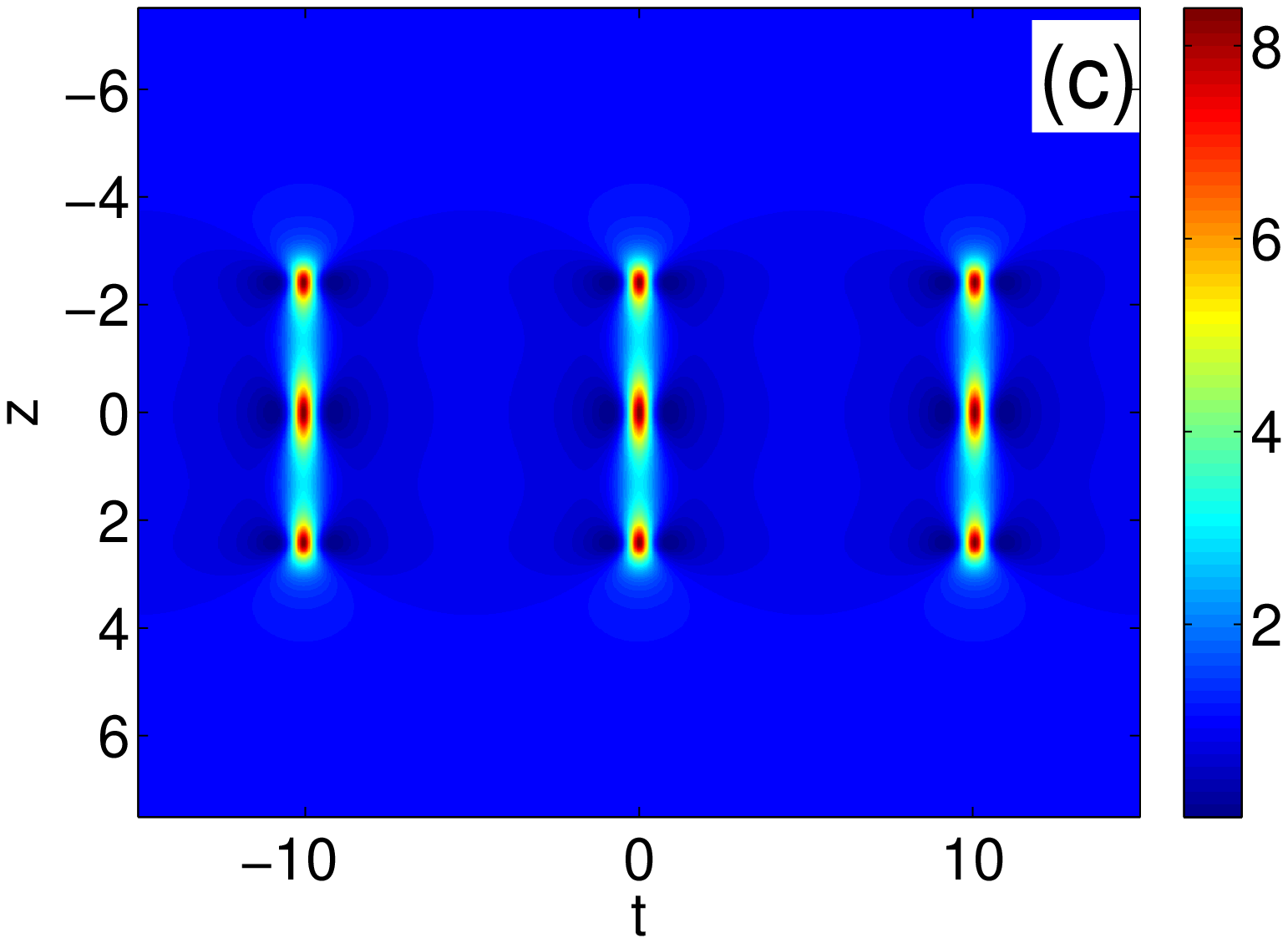}
\includegraphics[width=4cm,height=2cm]{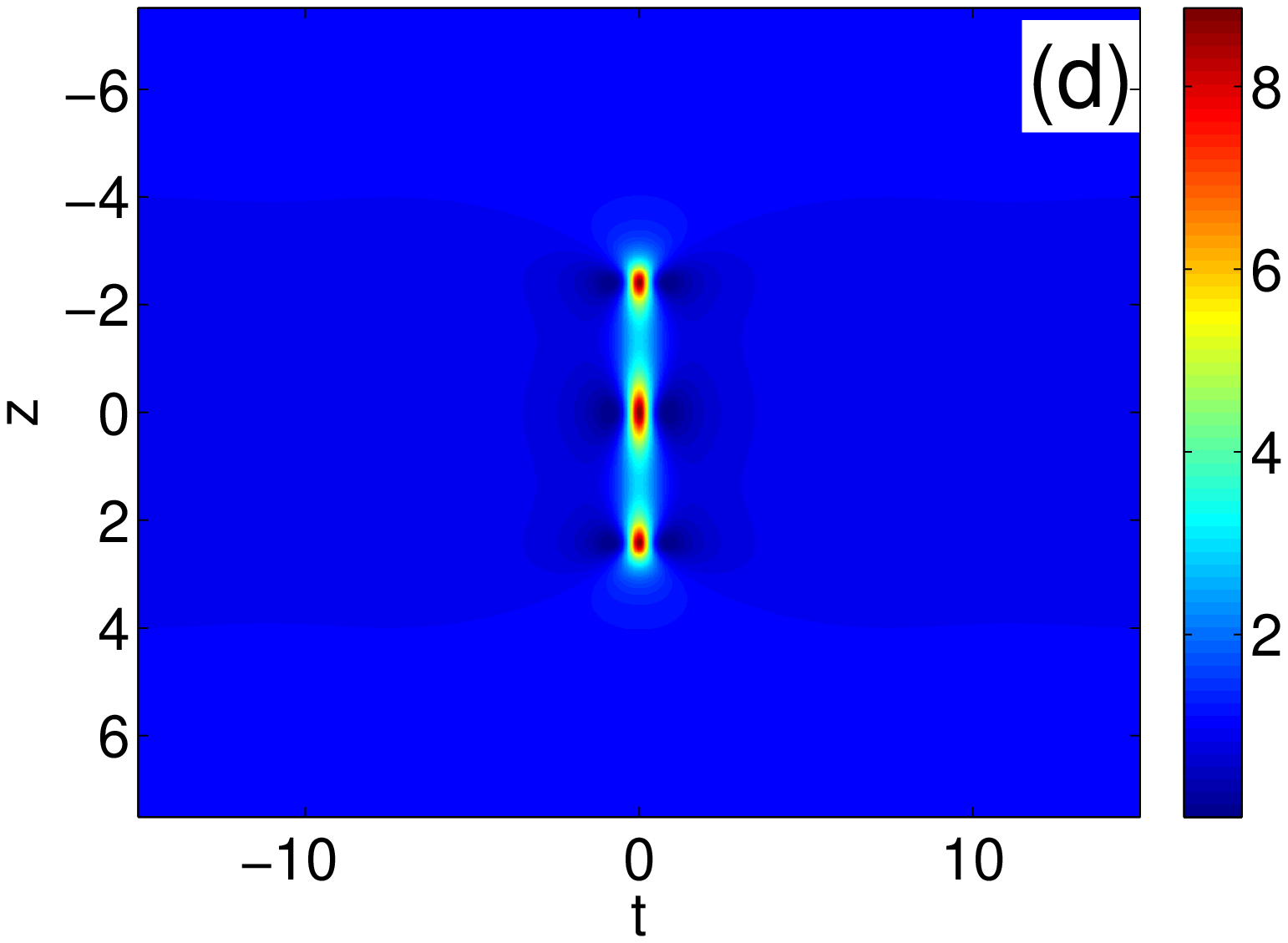}
\caption{ (Color online) Intensity of the solution~\eqref{eq:akhmed} showing the transition from generalized Akhmediev breather solutions to the Peregrine solutions as a function of $\eta$, (a) $\eta=0.65$, (b) $\eta=0.85$, (c) $\eta=0.95$, (d) $\eta=0.99$. The other parameters are
$d_m=2$, $A=1$, $z_0=0$, and $k_{m}=\pi/4$. \label{fig:1}
}\end{figure}

In this section, we shall introduce an explicit Peregrine-type solution of the above NLSE, in the presence of periodic modulations. It is well known that, to construct explicit solutions of the NLSE, one needs some integrability conditions~\cite{serkin2007_zhao}. Here, we shall work under the assumption that
 $RD_{2z}-D_{2}R_{z}=0$. This implies that $D_2(z)=cR(z)$ and through normalization we can always assume $c=-1$ since we work in the focusing case, where $D_2R\leq 0$. To construct the generalized Peregrine solution, we start with a seed solution in the form of a plane wave
\begin{equation}\label{eq:seed}
q_{0}(z)=A\exp[i\varphi_{0}(z)],
\end{equation}
which is the background solution from which the Peregrine-type rogue wave appears. The parameter $A$ determines the initial amplitude of the background, and
\begin{equation}
\varphi_{0}(z)=A^2\int_{z_0}^{z}
R(z')dz',
\end{equation}
where $z_0$ fixes a global phase.  From the seed $q_{0}$, one can first construct generalized inhomogeneous AB solutions  through the use of the Darboux transformation method 
\cite{guo}. For $0\leq \eta <A$, this yields
\begin{widetext}
\begin{equation}\label{eq:akhmed}
q_{\mathrm{AB}}=A\left[
1+\frac{2\eta\left(A\cos(d_{1})-\eta\cosh(d_{2})-i\alpha\sinh(d_{2})\right)}{A\cosh(d_{2})-\eta\cos(d_{1})}\right]\exp(i\varphi_{0}),
\end{equation}
\end{widetext}
with 
\begin{eqnarray*}
d_{1}(t)&=&2\alpha t,\\
d_{2}(z)&=&2\alpha\eta\int_{z_0}^{z} D_{2}(z) dz, \quad\alpha=\sqrt{A^{2}-\eta^{2}}.
\end{eqnarray*}
This generalized AB reduces to the standard form when the coefficients in Eq.~\eqref{1} are constant ($R(z)=1$, $D_2(z)=-1$)~\cite{Akhmediev_Park}. Notice that the intensity $|q_{AB}(z,t)|^2$ of this solution exhibits a periodic modulation in $t$ with period ${\pi}/{\sqrt{A^{2}-\eta^2}}$, which tends to $+\infty$ as $\eta\to A$.  
As a result, in the limit $\eta\rightarrow A$, the above generalized AB reduces to a generalized PS given by
\begin{widetext}
\begin{equation}\label{eq:PS}
q_{\mathrm{PS}}(z,t)=A\left[
 4\frac{1-2iA^{2}\int_{z_0}^{z}
D_2(z^{'})dz^{'}}{1+4A^{4}\left(
\int_{z_0}^{z}
D_2(z^{'})dz^{'}\right)
^{2}+4A^{2}t^{2}}-1\right]
e^{i\varphi_{0}(z)}.
\end{equation}
\end{widetext}
The same solution can also be obtained using a similarity transformation, as in \cite{Dai}.
In what follows we will study this solution in detail, and characterize its main properties when the periodic dispersion is of the form 
\begin{equation}\label{eq:D2}
D_2(z)=-1+d_m\cos(k_{m}z),
\end{equation}
where $d_m$ is the amplitude of modulation and $k_{m}$ its
spatial frequency. Note that in this analytical study $d_m$ can take arbitrary values since the method used is quite general and there is no need to assume small values of $d_m$. The spatio-temporal characteristics of the solution~\eqref{eq:akhmed} are illustrated in Fig.~\ref{fig:1} where the transition from the  generalized AB solutions~\eqref{eq:akhmed} to the generalized
PS solutions~\eqref{eq:PS} as $\eta\rightarrow A$ is shown for a modulation amplitude $d_m=2$ and with $z_0=0$. 
Note that both the AB and the limiting PS display multiple compression points, located at different values of $z$, a phenomenon that does not occur in the standard AB or PS of the NLSE with constant coefficients, which displays a unique compression point. This new feature of the generalized PS results from the longitudinal modulation in the coefficients in the NLSE and is strongly linked to the value of $d_m$, which determines the number of compression points that occurs in generalized PS, as is shown in the next section.
\begin{figure}[b]
\includegraphics[width=8cm,height=4cm]{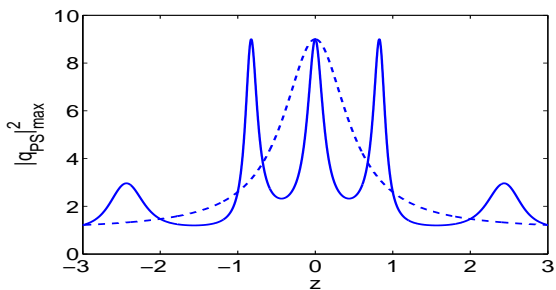}
\caption{(Color online) The shape of the generalized Peregrine Rogue Wave (see Eq.~\eqref{eq:max}) as a function of $z$, with $A=1$ and $k_m=\pi$ for $d_m=0$ (dashed line), and $d_m=5$  solid line. \label{fig:shapePRW}}
\end{figure} 
%%%%%%%%%%%%%%%%%%%%%%
\section{Multiple compression points: the Peregrine comb}
We now consider the generalized PS of Eq.~\eqref{eq:PS}, with $z_0=0$, and the periodic dispersion coefficient given by Eq.~\eqref{eq:D2}. At a fixed point $z$ along the fiber, the maximum value of the wave's intensity is obtained at $t=0$ and is given by
\begin{equation}\label{eq:max}
|q_{\mathrm{PS}}(z,0)|^2=|q_{\mathrm{PS}}|_{\max}^{2}(z)=\frac{9+40A^{4}Z
^{2}+16A^{8}Z^{4}}{(1+4A^{4}Z^{2})^{2}}A^{2},
\end{equation}
where we have set 
\begin{eqnarray}\label{eq:Z}
Z=z-\frac{d_m}{k_{m}}\sin(k_{m}z)=z\left(1-d_m\frac{\sin(k_m z)}{k_m z}\right).
\end{eqnarray}
The above expression provides a first interesting result. Indeed, this expression is easily seen to reach its maximal value $9A^2$ whenever $Z=0$. Hence the periodic variation of the 
coefficients in the NLSE has no effect on the maximum intensity of the rogue wave, which is identical to the one obtained in the case of constant coefficients. Note that a similar result has
been reported in \cite{Usama} where the NLSE has been considered in
presence of a linear potential. On the other hand, the equation $Z=0$ has, beyond the obvious solution $z=0$, several other solutions, provided $d_m>1$, as is readily seen from~\eqref{eq:Z}. We will refer to those as the compression points of the RW. In other words, when $d_m>1$, the amplitude $|q_{\mathrm{PS}}(z,0)|$ of the generalized PS has several absolute maxima as a function of $z$. For such large values of the modulation, the shape of the PS is drastically altered. This phenomenon is the most striking new feature of the generalized Peregrine solutions in~\eqref{eq:PS}. It will be analyzed in more detail below and can be observed  in the last panel of Figures~\ref{fig:1} and  Fig.~\ref{fig:shapePRW}.

To better understand the temporal evolution of this generalized PS as well as its spatial shape, it is convenient to rewrite it in the following  form:
\begin{eqnarray}\label{eq:pulseCW}
q_{\mathrm{PS}}(z,t)=
\frac{2}{W}\frac{1}{
1+\left(\frac{t}{W}\right)
^{2}}e^{i\varphi_{1}(z)}-Ae^{i\varphi_{0}(z)},
\end{eqnarray}
 with
\begin{eqnarray}
\varphi_{1}(z)=\varphi_{0}(z)+\tan^{-1}(-2A^{2}Z),
\end{eqnarray}
\begin{equation}\label{eq:width}
W(z)=\frac{\sqrt{1+4A^{4}Z^{2}
}}{2A}.
\end{equation}
In this form, it is seen that the generalized PS is a superposition of a CW and a
pulse with characteristics $W$, $2/W$, that correspond to its temporal width, and its amplitude, respectively. In the temporal domain, the pulse has a Lorentzian shape with width $W$, depending on $z$.  The shorter this pulse, the intenser it is, since its maximal amplitude is given by $2/W$. This shows the generalized PS has the typical signature of a RW that ``appears from nowhere and disappears without a trace''~\cite{Akhmediev2009}. 
Note that $W$ is a function of $Z$, which is therefore the variable that controls the spatio-temporal behaviour of the pulse, as well as of the full generalized PS.  When $d_m=0$, so that the NLSE has constant coefficients, the width $W$ has a unique minimum at $z=0$, which is also the unique absolute maximum of both the pulse and of the generalized PS amplitude $|q_{\mathrm{PS}}|$ itself.
As $d_m$ increases from $d_m=0$ to $d_m=1$, the pulse flattens at its top, until, as $d_m$ increases beyond the critical value $d_m=1$, it gradually develops two extra compression points close to the central one at $z=0$, which are symmetrically positioned around the latter. To understand what happens for larger values of $d_m$, we proceed as follows. Whenever $d_m>1$, there are several compression points, for each value of $z\not=0$ for which $Z=0$, corresponding to the solutions to
$$
\frac{\sin(k_mz)}{k_mz}=\frac1{d_m}.
$$ 
\begin{figure}[t]
\includegraphics[width=4cm, height=3cm]{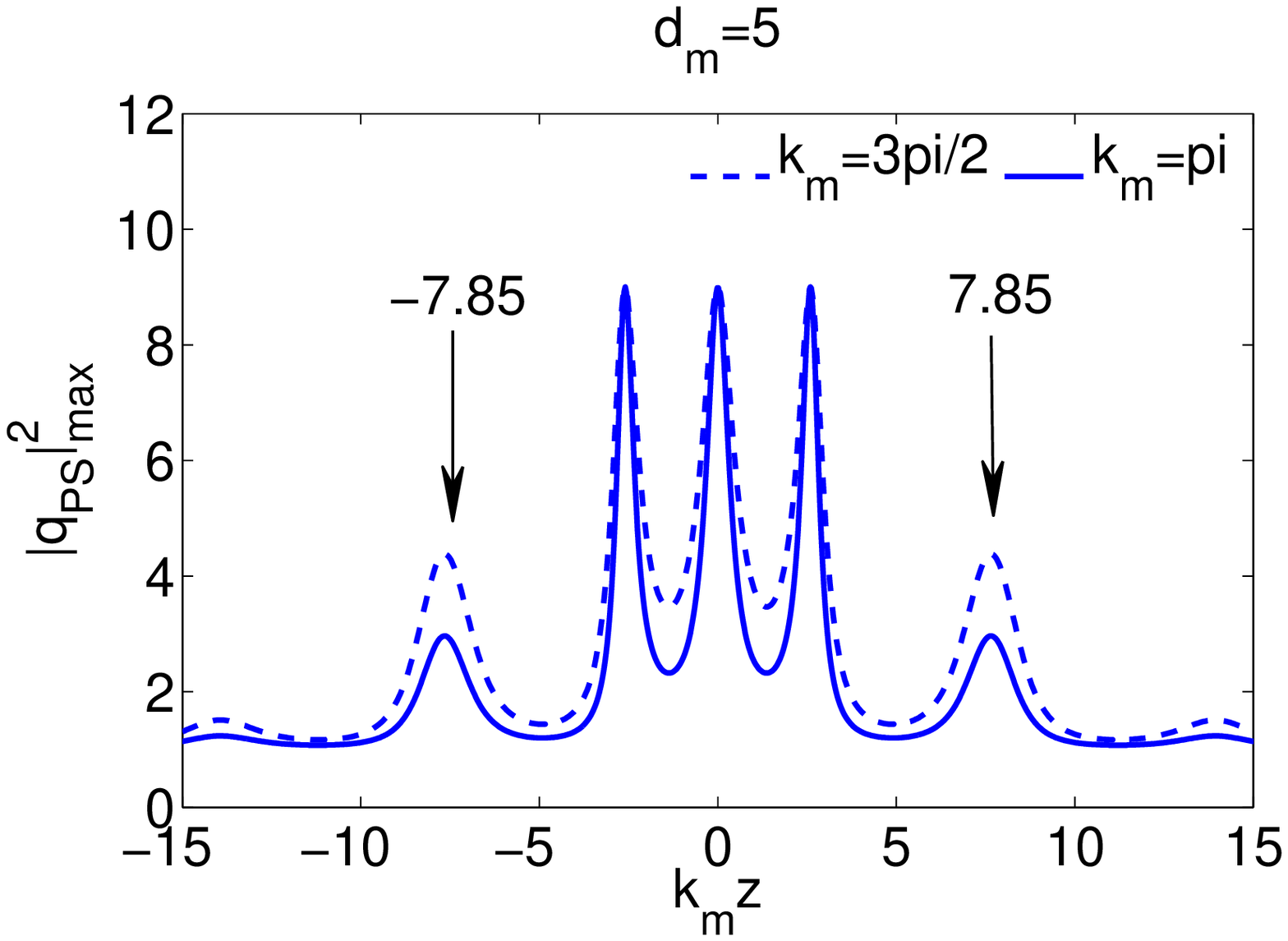}
\includegraphics[width=4cm, height=3cm]{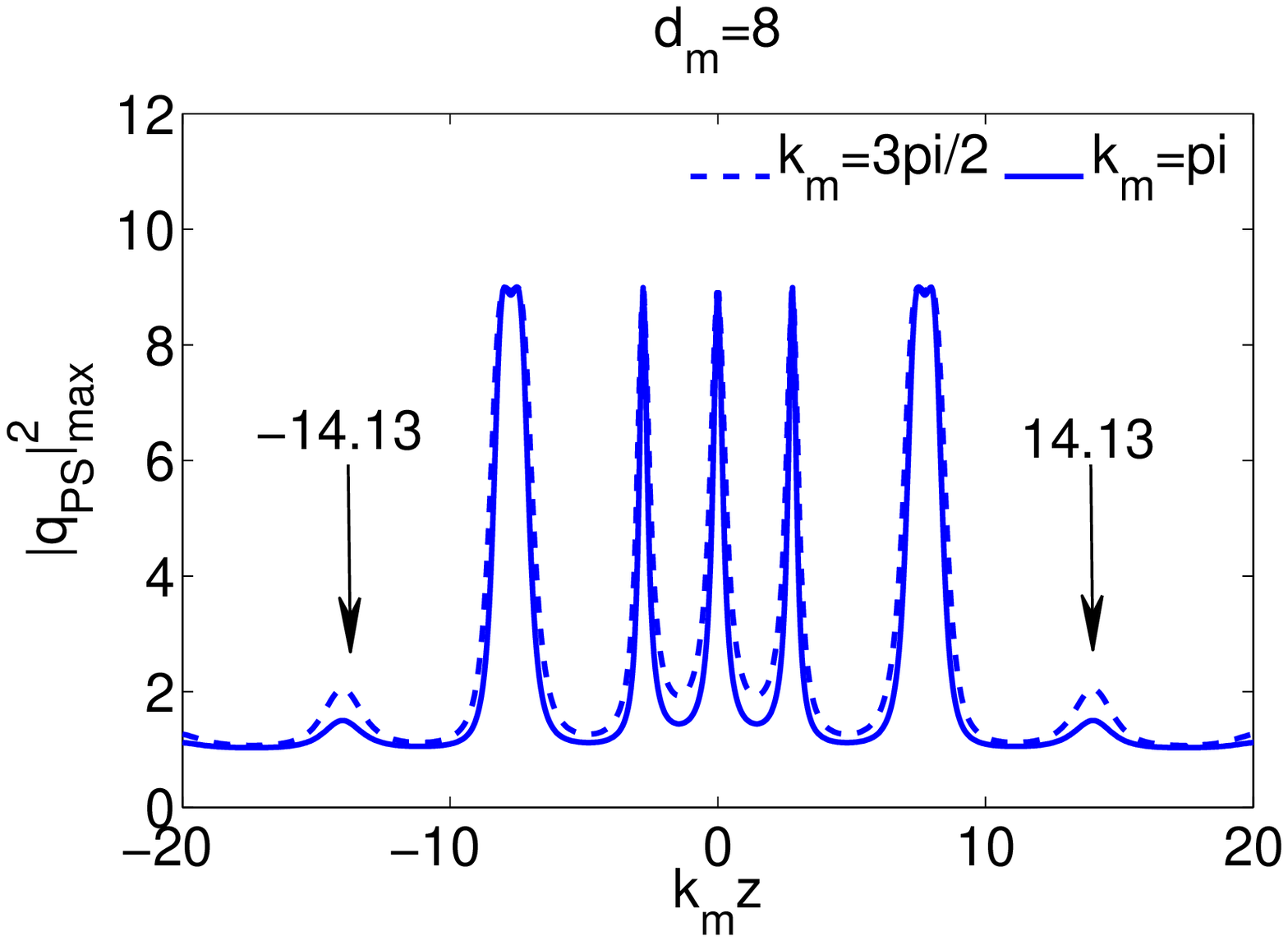}
\caption{ (Color online) Graph of $|q_{\mathrm{PS}}(z,0)|^2$ for values of $d_m$ and $k_m$ as indicated. Note the extra peaks close to $\pm k_mz= k_mz_1=d_{{\mathrm m}, 1}\simeq 7.85$ and to $\pm k_mz= k_mz_2=d_{{\mathrm m}, 2}\simeq 14.13$. \label{fig:peaksemerge}}
\end{figure}
Whereas an explicit closed form expression for the solutions to this equation is not available, inspecting the graph of the sinc functions, it is clear that, as $d_m$ is increased, the number of solutions increases. It is easily seen that the threshold values $d_{m, \ell}$ of $d_m$ for which an extra pair of compression points appears are well approximated by 
\begin{equation}\label{eq:dmcritical}
d_{m,\ell}=\frac{\pi}{2}+2\ell\pi,\quad \ell\geq 1,
\end{equation}
corresponding to  new compression points approximately positioned at
(see Fig.~\ref{fig:peaksemerge}),
\begin{equation}
z=\pm z_\ell=\pm\frac{d_{m, \ell}}{k_m}.
\end{equation}
\begin{figure}[b]
\includegraphics[width=8cm, height=4cm]{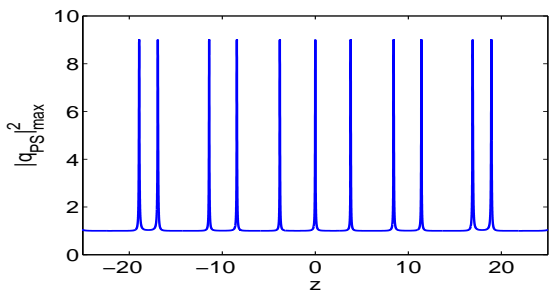}
\caption{ (Color online) The shape of the Peregrine comb for
$A=1$, $d_m=20$, and $k_m=\pi/4$
(see Eq. \eqref{eq:max})\label{fig:peregrinecomb}.}
\end{figure}
As $d_m$ increases from $d_{m,\ell}$ to $d_{m,\ell+1}$, the single compression point that formed close to $z_\ell$ splits in two, while a new smaller peak slowly emerges in the shape of the generalized PS, close to the value $z=z_{\ell+1}$. When $d_m$ reaches the value $d_{m,\ell+1}$, this new peak reaches the maximum value $9A^2$. This can be observed in Fig.~\ref{fig:peaksemerge}, for $d_m=8$, which is situated between $d_{m, 1}\simeq 7.85$ and $d_{m,2}\simeq 14.13$. One notices in the figure that the compression point at  $k_mz\simeq 7.85$ has started to split, whereas the local maximum at $k_mz\simeq 14.13$ is quite visible already.  Note that, for large $d_m$ the spacing between the successive peaks becomes approximately $\pi/k_m$, giving rise to a comblike structure with a periodicity of half the wavelength of the GVD modulation, that we shall refer to as the Peregrine comb (see Fig.~\ref{fig:peregrinecomb}). The width of the teeth of the comb scales as $d_m^{-1}$, as is readily checked.

\section{Further properties of the generalized Peregrine solitons}

\begin{figure}[b]
\includegraphics[width=8.0cm,height=4cm]{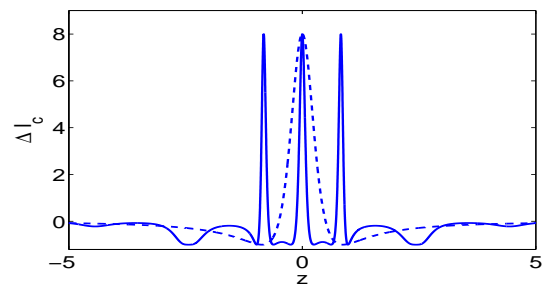}
\caption{(Color online) The distribution of the difference between
the light intensities of the PS and
the CW background at $t=0$ given by ~\eqref{eq:energy}.
Parameters $A=1$, $k_m=\pi$,
$d_m=0$ (dashed line), and $d_m=5$ (solid line).}
\label{fig:3}
\end{figure}

To shed further light on the spatio-temporal behaviour of the generalized PS, it is instructive to introduce the difference between the light intensity of the generalized PS and the CW background as follows \cite{Zhou, Yang}
\begin{eqnarray}\label{eq:energy}
\Delta I_{c}(t,z)&=& |q_{PS}(t,z)|^{2}-A^2\nonumber \\
 &=&8A^{2}\frac{1+4A^{4}Z^{2}-4A^{2}t^{2}}{\left[
1+4A^{4}Z^{2}+4A^{2}t^{2}\right]
^{2}}.
 \end{eqnarray}
It is easily checked that it possesses the property
$$
\int_{-\infty}^{+\infty}\Delta I_{c}(t,z)dt=0,
$$
 for all values of $z$. This is a reflection of the fact that the energy of the pump is preserved along the fiber, in spite of the periodic modulations in its physical characteristics.
It also implies that, at any value of $z$, the light intensity of the generalized PS will be sometimes higher, and sometimes lower than the background intensity (see Fig.~\ref{fig:3}).
In particular, at compression points, when $Z=0$, it will be lower for all $|t|\geq 1/2A$, to compensate for the very short burst of high intensity around $t=0$ (see Eq. \eqref{eq:energy}).

A related quantity, which is however not experimentally easily accessible,
is the energy of the Peregrine pulse~\cite{Zhou, Yang}:
\begin{eqnarray}\label{eq:exchange}
E_{\mathrm{pulse}}(z)&=&\int_{-\infty}^{+\infty}|q_{PS}(t,z)-q_{PS}(\pm\infty,z)|^{2}dt \nonumber\\
&=&\frac{4\pi A}{\sqrt{1+4A^{4}Z^{2}}}.
 \end{eqnarray}
One notices that the energy of the pulse is a simple function of the width $W$ and therefore clearly maximal at the compression points $Z=0$. 

\begin{figure}[b]
\includegraphics[width=.4\textwidth]{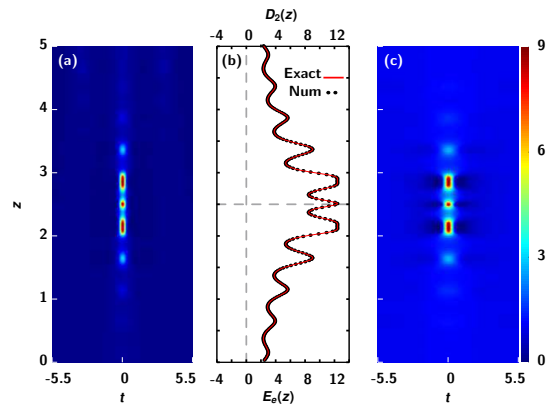}
\caption{ (Color online)  (a) Numerical solution of Eq.~\eqref{eq:nlse} with initial condition given by the generalized Peregrine
solution~\eqref{eq:PS} at $z=0$. (b) Pulse energy $E_{\mathrm{pulse}}(z)$. (c) analytical profile of the generalized Peregrine solution ~\eqref{eq:PS}.
The parameters are $z_0=2.5$, $d_m=5$, $k_{m}=4\pi$, and $A=1$.}
\label{fig:5}
\end{figure}

We expect the generalized PS to be unstable, which poses the question whether it can be obtained with a numerical integration of the NLSE. To check this, we have taken the generalized PS \eqref{eq:PS} as initial condition, and simulated its subsequent evolution by numerically solving Eq.~\eqref{eq:nlse}. The results are summarized in Fig.~\ref{fig:5}, which displays the intensity profile of the generalized PS obtained numerically (Fig.~\ref{fig:5}(a)) as well as the analytical and numerical pulse energy (red line and dotted black line, respectively, in Fig.~\ref{fig:5}(b)). The analytically computed intensity profile of the PS \eqref{eq:PS} is shown in Fig.~\ref{fig:5}(c). As can be seen in the figure, the numerically computed PS exhibits 3 compression points as expected for the value of $d_m=5$ used. In addition, the numerical intensity profile of Fig.~\ref{fig:5}(a) is in reasonable agreement with the analytical one in Fig.~\ref{fig:5}(c). Similarly, the numerical and analytical results displayed in Fig.~\ref{fig:5}(b) are in excellent agreement.%, in particular, for the locations of compression points (red line in Fig.~\ref{fig:5}(b)), although the numerics underestimates the analytical absolute maxima of the energy of the pulse. 

To end this section, we compute the frequency spectrum of the PS in~\eqref{eq:pulseCW}, which is the most accessible physical quantity in experiments:
\begin{widetext}
\begin{eqnarray}\label{eq:spectrum}
F(\omega,z)&=&\frac{1}{\sqrt{2\pi}}\int_{-\infty}^{\infty}q_{PS}(t,z)e^{i\omega
t}dt\nonumber \\
&=&\sqrt{2\pi}\left[\frac{1-2iA^{2}Z}{\sqrt{1+4A^{4}Z^{2}}}e^{{-\frac{1}{2A}|\omega| \sqrt{1+4A^{4}Z^{2}}}}-A\delta(\omega)\right]e^{i\varphi_{0}(z)}.
\end{eqnarray}
\end{widetext}

\begin{figure}[t]
\includegraphics[width=4.0cm,height=4cm]{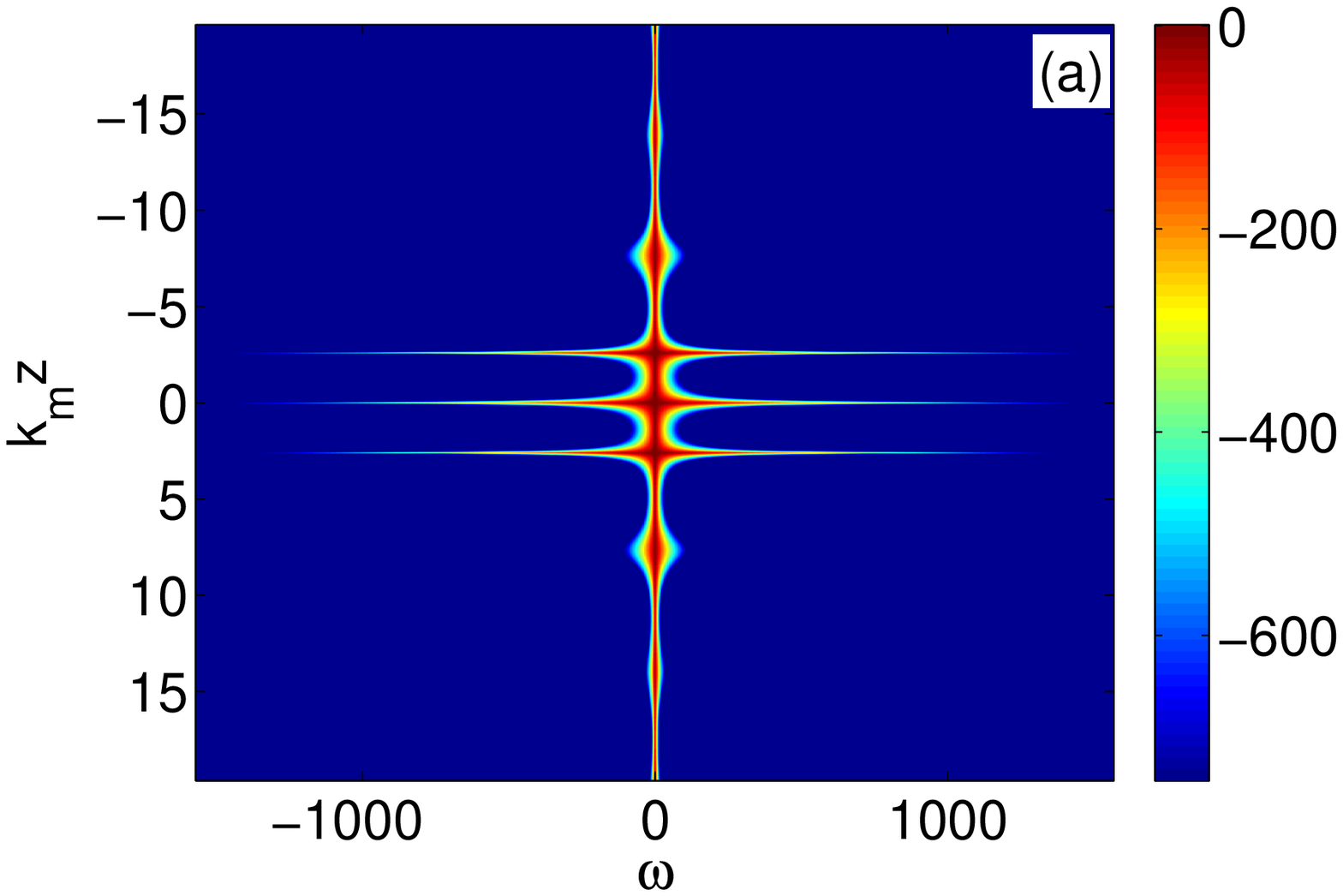}
\includegraphics[width=4.0cm,height=4cm]{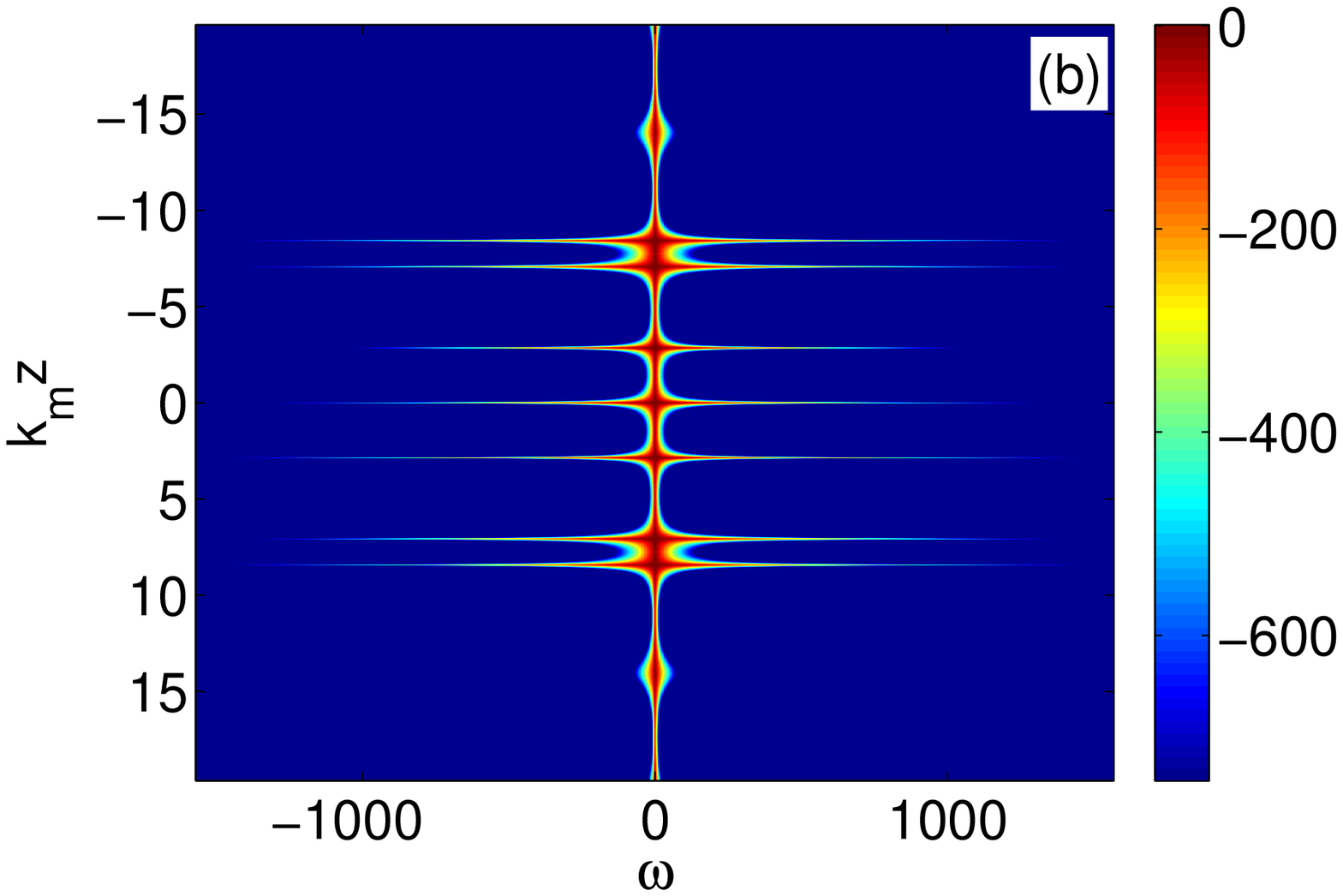}
\caption{(Color online) The spectrum of the generalized Peregrine soliton log scale, i.e, $\log|F(\omega,z)|$ with A=1. (a) $d_m=5$, (b) $d_m=10$.  
\label{fig:7}}
\end{figure}

The Dirac delta function $\delta(\omega)$ originates from the
finite background level.
The modulus of this spectrum with $A=1$ is given by:
\begin{equation}\label{eq:abs_spectrum}
|F(\omega,z)|=\sqrt{2\pi}e^{
{-0.5|\omega|\sqrt{1+4Z^{2}}}}.
\end{equation}
The Dirac delta function is omitted here, so ~\eqref{eq:abs_spectrum} represents the
spectrum of the variable part of the solution. In the homogeneous case, the spectrum starts
with narrow spectral components and then spreads into
a triangular-type shape \cite{Dudley2008}. In the inhomogeneous case, the rogue wave starts also with narrow spectral components, but spreads and shrinks during the propagation along the fiber, and eventually recovers the initial shape (see  Fig.~\ref{fig:7}(a)). As can be seen by comparing Fig.~\ref{fig:peaksemerge} and Fig.~\ref{fig:7}, each nonlinear spreading in the spectrum is associated with a corresponding maximal compression point. Thus, when the amplitude of modulation increases, the number of spectral components increases as shown in Fig. \ref{fig:7}(b).
 
\section{Beyond integrability}
\begin{figure}[b]
\includegraphics[width=9.0cm]{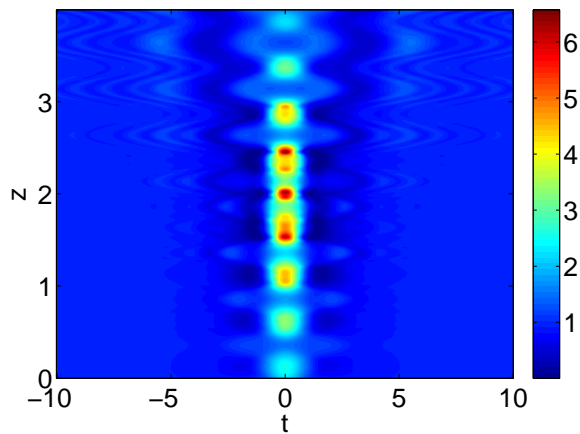}
\caption{(Color online) Numerical solution of Eq.~\eqref{eq:nlse} in the nonintegrable case with initial condition given by the generalized Peregrine solution~\eqref{eq:PS} at $z=0$ with $z_0=1.25$ and $A=1, d_m=6.6, k_m=4\pi$. 
\label{fig:8}}
\end{figure}

The analytical expression~\eqref{eq:PS} provides a solution of the NLSE only if the integrability condition is satisfied. However, this will rarely be the case in realistic situations, and in particular in optical fibers, where the nonlinearity coefficient $R(z)$ cannot change sign. In this section, we show that solutions with characteristics similar to those of the generalized PS can still be generated in such cases. For that purpose, we {first solve the dimensionless} Eq. ~\eqref{eq:nlse} by means of the split-step
Fourier method with the initial condition given by the PS ~\eqref{eq:PS} for $z_{0}=1.25$. The dispersion and nonlinear coefficients are choosen as $D_2(z)=-1+6.6\cos(4\pi z)$ and $R(z)=1$. Figure~\ref{fig:8}(a) shows the spatiotemporal evolution of the generalized Peregrine solution in the nonintegrable case and we observe that there still are multiple compression points. From the numerical profile of the generalized Peregrine solution, it can be seen that the number of peaks corresponding to the amplitude of modulation $d_m=6.6$ is equal to three, in agreement with our theoretical prediction.

\begin{figure}[t]
\includegraphics[width=8.750cm]{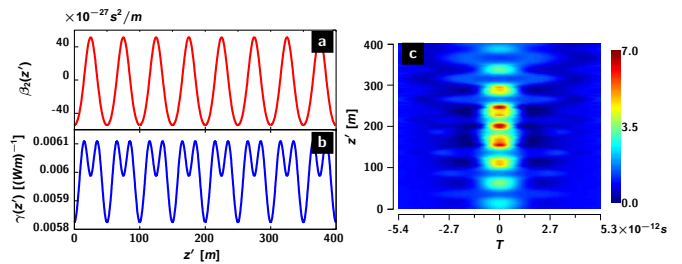}
\caption{(Color online)  Evolution plots in experimental case (a) dispersion, (b) nonlinearity, (c) intensity of Peregrine soliton type solution for the experimental parameter values as described in the text.}
\label{fig:6}
\end{figure}

As a further illustration, we have chosen for $D_2$ and $R$ values corresponding to those of an experimentally
realizable photonic crystal fiber with a GVD and nonlinear coefficients that are periodically modulated along the  {physical propagation axis $z'$ and the real time $T$. Notice that, the physical variables $z'$ and $T$ are related to the associated dimensionless quantities in Eq. ~\eqref{eq:nlse} as $z=z'/L_{NL}$ and $t=T/\sqrt{(|\langle\beta_2\rangle|L_{NL})}$ where the nonlinear characteristic length $L_{NL}$ is equal to $1/(\langle\gamma\rangle P_0)$, and $\gamma$ and $P_0$ are respectively the Kerr nonlinearity and the incident peak power.}    
We have assumed the air filling fraction $d/\Lambda=0.4$ ($d$ is the air hole diameter, $\Lambda$ is the hole pitch of the periodic cladding) to be constant so that by varying $\Lambda$, $d$ is adjusted accordingly. The
modulation amplitude corresponds to $\pm12\%$ of the average hole pitch ($\Lambda_{0}=2~\mu \mathrm{m}$). By using the above characteristics of the photonic crystal fiber, the periodic evolution of the GVD and the nonlinear coefficient vs the  longitudinal $z'$ axis was calculated from the model \cite{Saitoh} at the pump wavelength $\lambda_{p}=1700~\mathrm{nm}$ and pump power $P_0=1.7~\mathrm{W}$.
The results are shown in Figs. \ref{fig:6}(a) and (b). As seen in Fig.~\ref{fig:6}(a), the second-order dispersion $\beta_{2}(z')$ undergoes, when compared to its average value, large oscillations, equivalent to $d_m\simeq 6.6$ and $k_m=4\pi$ in Eq.~\eqref{eq:D2}.
The relative longitudinal variations of the nonlinear coefficient $\gamma(z')$  {shown in Fig. \ref{fig:6}(b)}
 are much lower ($\simeq 0.004\%$), so that the fiber can indeed be seen as a mainly dispersion-managed
device {for which the nonlinear coefficient $\gamma(z')$ can be taken to a good approximation to a constant. For more details about the experiment see ~\cite{Drokes1}.} In this realistic situation, the periodic coefficients $\beta_{2}(z')$ and $\gamma(z')$ do clearly not satisfy the integrability condition. 
We thus use the results obtained in Figs. \ref{fig:6}(a) and (b), and we integrate numerically the inhomogeneous NLSE Eq.~\eqref{eq:nlse} with the initial condition given by the generalized PS~\eqref{eq:PS} with $z_{0}=1.25${, expressed in the original physical variables given above}. The result is displayed in  Fig.~\ref{fig:6}(c) showing that the solution still generates multiple compression points. 

This result is of importance since it indicates that the generalized PRW is a robust solution that persists in situations where the integrability condition is not satisfied.  As a result, we expect that multiple compression points could be observed in nonlinear fibers with periodically modulated characteristics. Note that the standard (one compression point) PRW has been observed recently \cite{Solli2007,Dudley,Kibler}  in optical nonlinear fibers, operating in the anomalous dispersion regime with constant coefficients.
%%%%%%%%%%%%%%%%%%%%%%%%%%%%%%%%%%%%
\section{ Conclusion}

\noindent We have carried out a theoretical investigation of the generalized Peregrine soliton of a completely integrable nonlinear Schr\"odinger equation  with periodically varying coefficients.  The most striking feature of this solution is the existence of multiple compression points, corresponding to points of maximal energy concentration,  in contrast with the unique compression point occurring in the case with constant coefficients. The number of compression points depends on the modulation amplitude but not on the modulation frequency of the coefficients. The amplitude of the solution at its compression points is identical to the one in absence of modulation.  We have furthermore demonstrated numerically that solutions with similar characteristics exist even when the NLSE is not completely integrable and can occur also in physically realizable  photonic crystal fibers. 

\section{ Acknowledgements}
C. G. L. T. acknowledges the support of the "Laboratoire d'Excellence CEMPI: Centre Europ\'een pour les Math\'ematiques, la Physique et leurs Interactions'' (ANR-11-LABX-0007-01). This research was supported in part by the Interuniversity Attraction Poles program of the Belgium Science Policy Office under the grant IAPP7-35 and the French Agence Nationale de la Recherche project OptiRoc ANR-12-BS04-0011.

\end{document}